\documentclass[aps,prx,reprint,amsmath,amssymb,showpacs]{revtex4-1}

\usepackage{siunitx}
\usepackage{subfigure}
\usepackage{color}

\usepackage{graphicx}
\usepackage{dcolumn}
\usepackage{bm}
\usepackage{mathtools}
\usepackage{xfrac}
\usepackage{soul}

\newcommand{\InRefFig}[1]{Figure~\ref{#1}} 

\newcommand{\RefFig}[1]{Fig.~\ref{#1}}
\newcommand{\RefFigs}[1]{Figs.~\ref{#1}}

\newcommand{\RefEq}[1]{Eq.~(\ref{#1})}
\newcommand{\RefEqs}[1]{Eqs.~(\ref{#1})}

\newcommand{\vres}{{v_\mathrm{res}}}
\newcommand{\vthr}{{v_\mathrm{thr}}}
\newcommand{\vmin}{{v_\mathrm{min}}}
\newcommand{\taus}{{\tau_\mathrm{s}}}
\newcommand{\taud}{{\tau_\mathrm{d}}}

\begin{document}


\title{Equivalence between synaptic current dynamics and heterogeneous propagation delays in spiking neuron networks}

\author{Matteo~Biggio$^1$}
\author{Marco~Storace$^{1}$}
\author{Maurizio~Mattia$^{2}$}%
\email{maurizio.mattia@iss.it}
\affiliation{%
$^{1}$DITEN, University of Genoa, 16145 Genova, Italy
}%
\affiliation{%
$^{2}$ Istituto Superiore di Sanit\`{a}, 00161 Roma, Italy
}%

\date{\today}

\begin{abstract}
Message passing between components of a distributed physical system is non-instantaneous and contributes to determine the time scales of the emerging collective dynamics like an effective inertia. In biological neuron networks this inertia is due in part to local synaptic filtering of exchanged spikes, and in part to the distribution of the axonal transmission delays. How differently these two kinds of inertia affect the network dynamics is an open issue not yet addressed due to the difficulties in dealing with the non-Markovian nature of synaptic transmission. Here, we develop a mean-field dimensional reduction yielding to an effective Markovian dynamics of the population density of the neuronal membrane potential, valid under the hypothesis of small fluctuations of the synaptic current. The resulting theory allows us to prove the formal equivalence between local and distributed inertia, holding for any synaptic time scale, integrate-and-fire neuron model, spike emission regimes and for different network states even when the neuron number is finite.
\end{abstract}


\maketitle


\section{Introduction}

Large distributed systems like brain neuronal networks often have to satisfy both timing and space constraints irrespective of their size. Indeed, they have to be compact in order to be portable, and at the same time their components (i.e., the neurons) have to communicate always to the same pace, which in turn is determined by the environment \cite{Buzsaki2013}. The parallel non-instantaneous communication of messages like the spikes emitted by the neurons can be seen as an effective inertia and can be increased acting locally, by slowing down the dynamics of the single elements composing the networks, or globally, by distributing inertia across 
the system elements and their connections. A large local inertia implies a long memory (i.e., a long decay time bringing to forget initial conditions) and usually a big size, as in the case of a large capacitance in an electronic circuit or a large mass in a given physical system.
This aspect usually obstacles the miniaturization of a many-body system made up of elements (with a given memory capacity) with large local inertia. Indeed, the bigger the overall memory capacity, the larger the required space and/or the consumed energy per element \cite{Mead1989,Sarpeshkar1998,Laughlin2003}.
Alternatively, a global inertia can be implemented by delaying in time the interaction between the network components, thus implementing a distributed memory of the collective activity spread across the whole system, possibly hindering fast synchronizations and hence making the system slower and more stable 
\cite{Atay2003,Mattia2003}. 
Also this strategy has drawbacks, such as the implementation of delay lines with 
their related costs. For instance, a delayed distribution of the past activity/state of the system requires a memory resource or an additional energy demand in order to keep alive the messages to be exchanged between the network components \cite{Laughlin2003,Merolla2014}. 
Maybe it is not by chance that both these strategies coexist in a huge cellular network like the one composing our brain. Indeed, neuronal networks have both local low-pass filters at the level of synapses -- which select preferentially slow-varying information and provide local inertia -- and a global spread of their instantaneous activity through the axon and dendrites, which propagate emitted spikes with suited delays and provide global inertia. Then a natural question arises: are these two kinds of inertia affecting differently the network dynamics?

This is a challenging question. Indeed, synapses in neuron networks filter incoming spikes with a wide variety of time scales, affecting the stability of various collective dynamics \cite{Wang1999}, the selectivity in transmitting information \cite{Abbott2004} and the reactivity to suddenly appearing exogenous stimuli \cite{Fourcaud2002}. Even for extremely simplified models of spiking neurons and network connectivities, theoretical approaches including non-instantaneous transmission rely on approximations valid for relatively small time scales \cite{Brunel1998,Brunel2001,Fourcaud2002,Schuecker2015}, or for quasi-adiabatic dynamical regimes \cite{MorenoBote2004,Moreno-Bote2006a}, or for neurons working with low-noise supra-threshold inputs \cite{Lindner2004b,Schwalger2015}.
Despite the long history of attempts in statistical physics to work out an effective one-dimensional Markovian description of the dynamics of a non-Markovian system \cite{Risken1984,Hanggi1995} (i.e., a suitable representation of the membrane potential dynamics of the neurons when synaptic filtering is incorporated), a theoretical framework valid for any correlation time of the synaptic input and any neuronal activity regime is still lacking if the boundary conditions describing the spike emission process are taken into account. 

Here, we address this issue by developing a theoretical framework in which from small to large synaptic time scales the same dynamical model holds for the instantaneous firing rate of spiking neuron networks. 
Starting from the population density dynamics of single-neuron state variables under mean-field approximation, we extend to the colored-noise case the spectral expansion of the associated Fokker-Planck (FP) equation  previously derived under white-noise assumption \cite{Mattia2002}. Resorting to a kind of central moment closure method \cite{Levermore1996}, a zero-th order approximation of the population density dynamics is worked out, recovering the same theoretical description as for the white-noise case. In the new model, the equation coefficients are modulated by the low-pass filtered mean synaptic current. This dimensional reduction provides results showing a remarkable agreement with microscopic simulations and finally allows us to provide an answer to the challenging question asked above. This is done by proving a formal equivalence between local and distributed inertia: a suited distribution of spike transmission delays between neurons (global inertia) allows to fully reproduce the dynamical features of the same network having instead non-instantaneous synaptic transmission (local inertia). 
This equivalence is independent of integrate-and-fire (IF) 
neuron models (VIF, LIF, EIF), spike emission regimes (sub- and supra-threshold) and dynamical state (stable asynchronous fixed point and limit cycles) chosen for the network, even when finite-size fluctuations are taken into account.

\section{Local inertia in neuronal networks}

\begin{figure}[!ht]
\includegraphics[width=1.0\columnwidth]{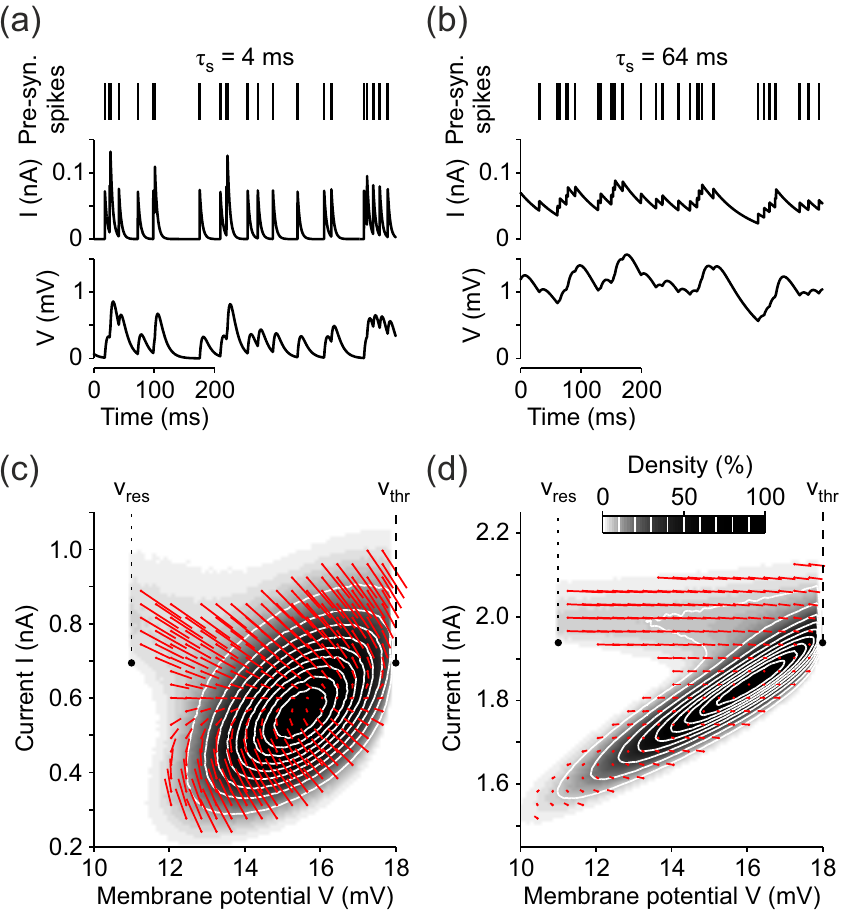} %
\caption{ %
Local inertia as spike input filtering by synaptic transmission. %
(a-b) Presynaptic activity modeled by a Poissonian spike train (top), is low pass-filtered by the synaptic transmission to produce the input current $I(t)$ (middle) to the membrane potential $V(t)$ (bottom) of a leaky integrate-and-fire (LIF) neuron. Examples are shown for fast (a, $\tau_s = 4$~ms) and slow (b, $\tau_s = 64$~ms) synaptic filtering. %
(c-d) Stationary distribution in the plane $(V,I)$ sampled from simulations of the same LIF neurons in (a-b) with slow (c) and fast (d) synaptic transmission. Red arrows: mean fluxes of realizations (probability currents) at different $(V,I)$. $\vthr$: absorbing barrier representing the spike emission threshold. $\vres$: reset membrane potential following the emission of a spike. Here $I_{ext} = 0$. %
}
\label{fig:TimeEvolutionOfVandI}
\end{figure}

As paradigmatic example of local inertia in a many-body distributed system, we use a network composed of $N$ integrate-and-fire (IF) neurons, each receiving an input current $I(t)$
resulting from the low-pass filtered linear combination of the spikes emitted at time $t_j$ by a subset of presynaptic cells in the network \cite{Destexhe1998,Brunel2001}:
$\tau_s \, \dot{I} = -I + J \, \sum_j \delta(t-t_j)$.
For simplicity, here we assume a first-order dynamics for synaptic current $I$ with decay time $\tau_s$ and constant efficacy $J$ (\RefFig{fig:TimeEvolutionOfVandI}a-b).
The membrane potential $V(t)$ of an IF neuron evolves according to the general equation
$\tau_m \, \dot{V} = - f(V) + R \, (I + I_{ext})$,
with membrane decay constant $\tau_m$, neuron resistance $R$ (henceforth set to 1 for simplicity) and a voltage-dependent leakage drift $f(V)$ which depends on the model neuron type. For instance, $f(V)$ is a constant drift for the perfect IF (PIF) neuron and $f(V) = V$ for the leaky IF (LIF)  \cite{Burkitt2006}. Neurons may receive an additional current $I_{ext}$ modeling external sources, like incoming synaptic input from other networks. Once $V$ crosses a threshold value $\vthr$, a spike is emitted and potential $V$ is reset to $\vres < \vthr$. Due to these boundary conditions, even under stationary regimes single-neuron dynamics is not analytically tractable. This is particularly apparent looking at the probability density $P(V,I)$ of states $(V,I)$ from a long time series  worked out numerically by modeling presynaptic activity as a Poissonian spike train (\RefFig{fig:TimeEvolutionOfVandI}c). The absorbing barrier in $\vthr$ and the reentering flux of realizations in $\vres$ modeling the emission of spikes, make $P(V,I)$ asymmetric \cite{Brunel1998,Fourcaud2002,Apfaltrer2006} 
and correlation between $V$ and $I$ non-monotonic, a feature even more apparent for relatively slow synaptic filtering (\RefFig{fig:TimeEvolutionOfVandI}d). %
Under diffusion approximation, holding for large rate of incoming spikes each only mildly affecting $V$ \cite{Tuckwell1988,Burkitt2006}, membrane potential dynamics is described by the following system of Langevin equations \cite{Risken1984}
\begin{equation}
   \dot{V} = -f(V)/\tau_m + I + I_{ext}\, , \quad
   \tau_s \, \dot{I} = -I + \mu_I + \sigma_I \, \xi \, ,
\label{eq:2DLangevin}
\end{equation}
where for simplicity currents are scaled by a factor $\tau_m$ (i.e., measured as voltage per unit time) and $I_{ext} = \mu_{ext} + \sigma_{ext} \, \xi_{ext}$ is a Gaussian white noise homogeneous in time.

When the mean driving
force alone (i.e., the deterministic part of the input $I +
I_{ext}$) is not enough to make the membrane potential $V$
cross the threshold $v_{thr}$, the neurons are evolving in a
\emph{noise-dominated} (or subthreshold) \emph{regime} and
irregular firing occurs, due to the stochastic part of the
input. In the opposite case, the emission of an action
potential can occur also in the absence of noisy afferent
currents, and the neurons are in a \emph{drift-dominated} (or
suprathreshold) \emph{regime} of activity, characterized by the
regular emission of spike trains \cite{Fusi1999,Brunel2000,Lansky2008}.

Equation~\eqref{eq:2DLangevin} describes the time course of a
generic realization $(V,I)$, which under mean-field
approximation represents the dynamics of the statistically
identical neurons of the network. According to this theoretical
ansatz, infinitesimal moments
\begin{equation}
\begin{array}{r}
\mu_I(t) = J \, C \, \nu(t)\\
\sigma_I^2(t) = J^2 \, C \, \nu(t)
\end{array}
\label{eq:moments}
\end{equation}
\noindent characterize the white noise $\xi$ driving the activity-dependent synaptic current \cite{Amit1997} as a function of the network emission rate $\nu(t)$ (i.e., the average number of spikes each neuron emits per time unit) and the average number of presynaptic contacts $C$ ($\leq N$) per neuron. The probability density $P(x,y,t)$ to find neurons at time $t$ with membrane potential $V(t) = x$ and synaptic current $I(t) = y$ follows the FP equation \cite{Risken1984}
\begin{equation}
\begin{array}{rcl}
	\partial_t P
	& = & - \partial_x \mathcal{F}_x(x,y) - \partial_y \mathcal{F}_y(y,\nu) \\
	& \equiv & \frac{1}{\tau_m} \, \partial_x \left[\left( f(x)-y-\mu_{ext}\right) +
         \frac{1}{2}\sigma_{ext}^2 \partial_x \right] P + \\
	&   & \frac{1}{\tau_s} \, \partial_y \left[(y-\mu_I) + \frac{1}{2} \sigma_I^2
         \partial_y \right] P \\
   & \equiv  & \left(\mathcal{L}_x + \mathcal{L}_y\right) P \, ,
\end{array}
\label{eq:2DFokkerPlanck}
\end{equation}
a continuity equation in which the density changes are determined by the divergence of the probability current $(\mathcal{F}_x,\mathcal{F}_y)$ (i.e., the flux of realizations) together with few specific boundary conditions (see \RefFig{fig:TimeEvolutionOfVandI}): i) spike emission as an absorbing barrier at $x = \vthr$ (only for $y>0$ if $\sigma_{ext}=0$), ii) reentering flux of absorbed realizations at the reset potential $\vres$ and iii) lower bound for the membrane potential implemented as a reflecting barrier at $\vmin$ \cite{Knight1996,Brunel1998,Brunel1999,Mattia2002}.

In general, \RefEq{eq:2DFokkerPlanck} is hard to solve and only approximated solutions (mainly under stationary conditions) have been worked out \cite{Risken1984,Doering1987,Hanggi1995,Fourcaud2002}. 
This is both due to the peculiar boundary conditions and to its intrinsic nonlinearity, as $\nu(t)$ depends on $P(x,y,t)$, being the flux of realizations crossing $\vthr$:
\begin{equation}
\nu(t) = \int_{-\infty}^\infty \mathcal{F}_x(\vthr,y) dy \equiv \int_{-\infty}^{\infty} \nu(y,t) dy \, ,
\label{eq:TotalEmissionRate}
\end{equation}
where $\nu(y,t) = \mathcal{F}_x(\vthr,y)$ is the `partial' rate of spikes emitted by subpopulation $y$. Here we derive a solution for this two-dimensional problem without making any assumption like stationarity in time and restricted ranges for the synaptic filtering timescale. We extend the spectral expansion approach exploited to study the one-dimensional case \cite{Knight1996,Mattia2002}, by describing $P(x,y,t)$ as the superposition of an \emph{infinite set of one-dimensional interacting sub-populations}, labeled by $y$. Sub-population dynamics are studied, as in \cite{Mattia2002}, by projecting $P$ at fixed $y$ on the eigenfunctions $\{|\phi_{m}\rangle\}$ of a suited one-dimensional FP operator, which here we set to be %
$\mathcal{L}_{xy} = \frac{1}{\tau_m}\partial_x [f(x)-y-\mu_{ext}] + \frac{1}{2\tau_m} \left[\sigma^2_{ext} + \sigma^2(y)\right] \partial_x^2$, %
with an additional diffusion term $\sigma^2(y)$ with respect to $\mathcal{L}_x$ of \RefEq{eq:2DFokkerPlanck}. This allows to rewrite this equation in terms of the expansion coefficients $a_{n}(y,t) = \langle \psi_{n} | P \rangle = \int_{\vmin}^\vthr \psi_{n}(x,y)\,P(x,y,t)\,dx$, which now explicitly depend on the auxiliary variable $y$ and where $\langle \psi_{n}|$ is the $n$-th eigenfunction of the adjoint operator  of $\mathcal{L}_{xy}$. The stationary mode with eigenvalue $\lambda_0 = 0$ has $\langle \psi_{0}| = 1$, thus $a_{0}(y,t)$ is the marginal probability density $\rho_{y}(y,t) = \int_{\vmin}^\vthr P(x,y,t) \, dx$.

The dynamics of $a_{n}$ can be worked out by deriving $a_{n}$ with respect to time \cite{Knight1996,Mattia2002}, and rewriting $\mathcal{L}_x = \mathcal{L}_{xy} - \frac{1}{2\tau_m} \sigma^2(y) \partial_x^2$:
\begin{equation}
\begin{array}{rcl}
	\dot{a}_{n} = \lambda_{n} a_{n} & + & \displaystyle \sum_{m}\langle \psi_{n}| \mathcal{L}_y \, a_{m}\phi_{m}\rangle \\
   & - & \displaystyle \frac{\sigma^2(y)}{2\tau_m}  \sum_{m} a_{m} \langle \psi_{n} | \partial_x^2 \phi_{m} \rangle \, .
\end{array}
\label{eq:RawCoeffDyn}
\end{equation}
For $n=0$, the dynamics of $\rho_{y}(y,t) = a_{0}(y,t)$ can be worked out, as both $\mathcal{L}_y$ and $a_n$ do not depend on $x$:
\begin{displaymath}
	\dot{\rho}_{y} = \mathcal{L}_y \, \rho_y - \rho_y \frac{\sigma^2(y)}{2\tau_m}
   \int_{\vmin}^\vthr \partial_x^2 \phi_0(x,y) \, dx \, .
\end{displaymath}
The integral on the r.h.s. can be solved by parts and turns out to be 0 thanks to both the conservation of the flux of realizations exiting from $\vthr$ and re-entering in $\vres$ ($\partial_x \phi_{0}|_{x=\vthr} = \partial_x \phi_0|_{x=\vres^+} - \partial_x \phi_0|_{x=\vres^-}$) and the reflecting barrier condition in $\vmin$, provided that the reasonable assumption $\phi_0(\vmin,y) = 0$ holds (this is surely true for $\vmin \to -\infty$, as $\phi_0$ is a probability density). Thus, the $\rho_y$ dynamics reduces to
\begin{equation}
	\dot{\rho}_y = \mathcal{L}_y \, \rho_y = \partial_y \frac{(y-\mu_I) \rho_y}{\tau_s} + \frac{\sigma_I^2}{2\tau_s} \, \partial_y^2 \rho_y \, .
\label{eq:MarginalDensity}
\end{equation}

Relying on the same spectral expansion as in \cite{Mattia2002}, each `partial' rate $\nu(y,t)$ in \RefEq{eq:TotalEmissionRate} can be decomposed as follows:
\begin{equation}
	\nu(y,t) = \rho_y(y,t) \, \Phi(y,\nu) + \sum_{n\neq 0} a_{n}(y,t) \, f_{n}(y,\nu) \, ,
\label{eq:PartialEmissionRate}
\end{equation}
where the stationary mode contribution is separated from the others. Here, $\Phi(y,\nu) \equiv \mathcal{F}_x(\vthr,y)|_{P = \phi_0}$ is the emission rate of the stationary one-dimensional density $\phi_0$ at fixed $y$, while the other nonstationary modes contribute with the fluxes $f_{n}(y,\nu) \equiv \mathcal{F}_x(\vthr,y)|_{P = \phi_n}$. Both $\Phi$ and $f_n$ depend on the total emission rate $\nu(t)$ through the input current moments $\mu_I$ and $\sigma_I$.

\section{Dimensional reduction of the emission rate dynamics}

Equations (\ref{eq:TotalEmissionRate}-\ref{eq:PartialEmissionRate}) together are a mere reformulation of the original problem \eqref{eq:2DFokkerPlanck}, not a solution. However, in this framework a perturbative approach can be envisaged. From \RefEq{eq:MarginalDensity}, synaptic current $y$ results to have time-dependent mean $\mu_y(t) = \langle y \rangle$ and variance $\sigma_y^2(t) = \langle \left[ y - \mu_y(t) \right]^2 \rangle$
\begin{equation}
\tau_s \, \dot{\mu}_y = -\mu_y+\mu_I \, , \qquad
\tau_s \, \dot{\sigma_y^2} = -2 \sigma_y^2 + \sigma_I^2 \, ,
\label{eq:CurrentMeanAndVar}
\end{equation}
like an Ornstein-Uhlenbeck process with nonstationary input moments $\mu_I(t)$ and $\sigma_I(t)$ \cite{Risken1984}. Hence, synaptic current displacement in time is $|y-\mu_y| = \mathcal{O}(\sigma_I)$ and, for small $\sigma_I$, we can assume a negligible role for a subpopulation labeled with $y$ distant enough from $\mu_y$. This assumption of a narrow $\rho_y$ centered around $y = \mu_y$ allows to simplify the integrals across the $y$ domain weighted by the expansion coefficients $a_n(y,t)$. Indeed, approximations of these integrals can be worked out by expanding in Taylor's series the $y$-dependent functions in the integrands. By combining Eqs. \eqref{eq:TotalEmissionRate} and \eqref{eq:PartialEmissionRate}, the total emission rate can be then rewritten as
\begin{equation}
\begin{array}{rcl}
\nu & = & \displaystyle \int_{-\infty}^\infty \nu(y,t) \, dy \\
    & = & \displaystyle \Phi(\mu_y,\nu) \, \int_{-\infty}^\infty \rho_y(y,t) \, dy \, + \\
		&   & \displaystyle \sum_{n\neq 0} f_{n}(\mu_y,\nu) \, \int_{-\infty}^\infty a_{n}(y,t) \, dy + \mathcal{O}(\sigma_I) \\
		& \equiv & \displaystyle \Phi_0(\nu) + \sum_{n\neq 0} a_{0n} \, f_{0n}(\nu) + \mathcal{O}(\sigma_I)  \, ,
\end{array}
\label{eq:ApproxTotalNu}
\end{equation}
where the $0$-th order terms of the Taylor expansions are defined as $\Phi_0(\nu) \equiv \Phi(\mu_y,\nu)$ and  $f_{0n}(\nu) \equiv f_n(\mu_y,\nu)$, together with the introduction of the new integral expansion coefficients $a_{0n}(t) \equiv \int_{-\infty}^\infty a_{n}(y,t) \, dy$. The next step to find a self-consistent approximated spectral expansion of \RefEq{eq:2DFokkerPlanck} is to work out the dynamics of these $a_{0n}(t)$. This can be done by integrating both sides of \RefEq{eq:RawCoeffDyn} and constraining the additional diffusion term in $\mathcal{L}_{xy}$ to be as small as the fluctuation size of $y$: $\sigma^2(y) = \mathcal{O}(\sigma_I^2)$. The dynamics of the integral expansion coefficients reduces to
\begin{displaymath}
\begin{array}{rcl}
	 \dot{a}_{0n} & = & \displaystyle \int_{-\infty}^\infty \lambda_n(y) \, a_n(y,t) \, dy \, + \\
	& & \displaystyle \sum_{m} \int_{-\infty}^\infty \langle \psi_{n}| \partial_y(\frac{y-\mu_I}{\tau_s}a_{m}\phi_{m})\rangle \, dy + \mathcal{O}(\sigma_I^2) \\
	& = & \lambda_{0n} \, a_{0n} - \displaystyle \frac{\mu_y-\mu_I}{\tau_s} \sum_{m} \langle \partial_y \psi_{n}| \phi_{m}\rangle|_{y=\mu_y} a_{0m} + \mathcal{O}(\sigma_I) \, ,
\end{array}
\end{displaymath}
where we integrated by parts the second integral and set $\lambda_{0n} \equiv \lambda_n(\mu_y)$. This equation for $a_{0n}$ can be further recast as follows, by taking into account \RefEq{eq:CurrentMeanAndVar}:
\begin{equation}
	 \dot{a}_{0n} = \lambda_{0n} \, a_{0n} + \displaystyle \dot{\mu}_y \sum_{m} \langle \partial_y \psi_{n}| \phi_{m}\rangle|_{y=\mu_y} a_{0m} + \mathcal{O}(\sigma_I) \, .
\label{eq:IntegrCoeffDyn}
\end{equation}
This is a dimensional reduction of \RefEq{eq:RawCoeffDyn} as the coefficients $a_{0n}$ do not depend on $y$, and it holds provided that current $I$ has a narrow distribution across neurons at each time $t$ (i.e., small $\sigma_I(t)$), leaving $\tau_s$ unconstrained.

To work out a self-consistent equation for emission rate $\nu(t)$, we still need to choose a suited $\sigma(y)$ in $\mathcal{L}_{xy}$. To this purpose, the case of instantaneous synaptic transmission $\tau_s = 0$ can be used as reference, as \RefEq{eq:2DFokkerPlanck} reduces to a one-dimensional FP equation with the only operator $\mathcal{L}_{x0} = -1/\tau_m \, \partial_x \left[\mu_I + \mu_{ext} - f(x)\right] + \left(\sigma_{ext}^2 + \sigma_I^2\right)/(2\tau_m) \partial_x^2$ \cite{Brunel1999,Fusi1999,Mattia2002}. Hence, $\mathcal{L}_{xy}|_{y=\mu_y}$ must tend to $\mathcal{L}_{x0}$ for vanishing $\tau_s$, which is the case if $\sigma^2(y) = J \, y$. Indeed, in this limit $\sigma^2(\mu_y) = J^2 \, C \, \nu = \sigma_I^2$, being $\mu_y = \mu_I = J \, C \, \nu$ from \RefEq{eq:CurrentMeanAndVar}. This implies that eigenvalues $\lambda_{0n}$ and eigenfunctions in \RefEq{eq:IntegrCoeffDyn} are the same as in the one-dimensional case with $\delta$-correlated synaptic input, but with collective emission rate $\nu(t) = \mu_y(t) / (J \, C)$.

In conclusion, this dimensional reduction extends the emission rate equation previously worked out for $\tau_s = 0$ \cite{Mattia2002}, and, by combining Eqs.~(\ref{eq:CurrentMeanAndVar}--\ref{eq:IntegrCoeffDyn}) and neglecting $\mathcal{O}(\sigma_I)$ terms, it results to be
\begin{equation}
	\begin{cases}
		\dot{\vec{a}}_0 = \left(\mathbf{\Lambda}_{0} + \mathbf{W}_{0} \, \dot{\mu}_y \right) \, \vec{a}_0 + \vec{w}_0 \, \dot{\mu}_y \\
		\dot{\mu}_y = \displaystyle \left(\mu_I(\nu) - \mu_y\right)/\tau_s \\
		\nu = \Phi_{0} + \vec{f}_0 \cdot \vec{a}_0
	\end{cases} \, .
\label{eq:LowDimERE}
\end{equation}
Here, the infinite vectors $\vec{a}_0 = \{a_{0n}\}$ and $\vec{f}_0(\nu) = \{f_{0n}(\nu)\}$ are introduced together with the eigenvalue matrix $\mathbf{\Lambda}_0 = \operatorname{diag}(\lambda_{0n})$. Separating stationary from non-stationary modes, the synaptic coupling vector $\vec{w}_0 = \{\langle \partial_y \psi_{n}| \phi_0\rangle|_{y=\mu_y}\}$ and matrix $\mathbf{W}_0 = \{\langle \partial_y \psi_n| \phi_m\rangle|_{y=\mu_y}\}_{m \neq 0}$ are also included, respectively.
For the above chosen $\sigma(y)$, we remark that all these terms depend on the total emission rate $\nu(t)$ only through $\mu_y(t)$.

\section{Local and distributed inertia equivalence}\label{sec:localvsdistrib}

So far, we considered the filtering activity operated by local synapses transmitting incoming spikes as a post-synaptic potential non-instantaneous in time. Here, we recall the known dynamics of a network of spiking neurons where synaptic transmission is instantaneous but a distribution of axonal delays is taken into account \cite{Brunel1999,Mattia2002,Mattia2003}. In this one-dimensional case ($\tau_s = 0$), the network dynamics can be simply worked out by replacing $\nu$ in $\mu_I$ and $\sigma_I^2$ (see \RefEqs{eq:moments}) with the instantaneous rate $\tilde{\nu}(t) = \int_0^\infty \nu(t-\delta) \rho(\delta) \mathrm{d}\delta$ of spikes received by neurons when a distribution $\rho(\delta)$ of axonal transmission delays $\delta$ is taken into account. Interestingly, the coefficients of the related emission rate equation result to have a straightforward relationship with those in Eq.~\eqref{eq:LowDimERE}. Indeed, the elements of the synaptic coupling vector and matrix now are $\langle \partial_{\tilde{\nu}} \psi_{n}| \phi_m\rangle = J \, C \langle \partial_y \psi_{n}| \phi_m\rangle|_{y=\mu_y}$ for any $n$ and $m$, since
\begin{displaymath}
	\tilde{\nu} = \frac{\mu_y}{J\,C}
\end{displaymath}
for what shown above. Due to this, the searched emission rate equation reduces to
\begin{equation}
	\begin{cases}
		\dot{\vec{a}} = \left(\mathbf{\Lambda}_{0} + \mathbf{W}_{0} \, J \, C \, \dot{\tilde{\nu}} \right) \, \vec{a} + \vec{w}_0 \, J \, C \, \dot{\tilde{\nu}} \\
		\dot{\tilde{\nu}} = \displaystyle \left(\nu - \tilde{\nu}\right)/\tau_\delta \\
		\nu = \Phi_{0} + \vec{f}_0 \cdot \vec{a}
	\end{cases} \, ,
\label{eq:DelayDistribERE}
\end{equation}
where the specific delay distribution
\begin{equation}
	\rho(\delta) = \frac{1}{\taud} e^{-\delta/\taud} \, \Theta(\delta)
\label{eq:ExpDelayDistrib}
\end{equation}
has been taken into account. This leads $\tilde{\nu}$ to be a version of the collective emission rate $\nu$ smoothed in time by a first-order low-pass filter with decay time $\tau_\delta$. $\Theta(\delta)$ is the Heaviside function as only positive transmission delays are admitted.

\begin{figure}[!ht]
\includegraphics[width=1.0\columnwidth]{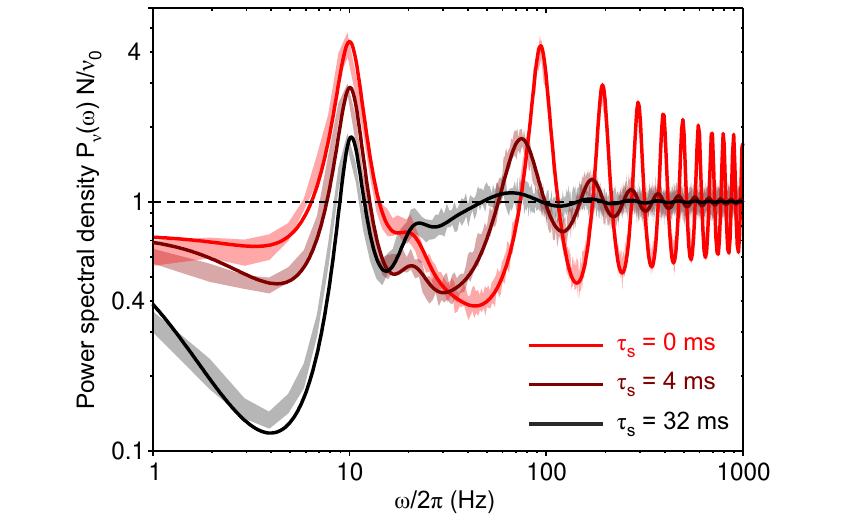} %
\caption{ %
Match between theory and simulations in networks of VIF neurons with different synaptic decay times $\taus$, under a quasi-stationary asynchronous state. Power spectral densities $P_\nu(\omega)$ of the network emission rate $\nu(t)$ are shown for $\taus = \{0,4,32\}$~ms. All networks have a stable fixed-point at $\nu^* = 10$~Hz, and are composed of $N = 2000$ excitatory VIF neurons. Synaptic matrices are random with connection probability $C/N = 5\%$ and average synaptic efficacy $J = 7.5 \, 10^{-3} \, \vthr$. Spikes are delivered to the postsynaptic targets with a fixed transmission delay $\delta = 10$~ms (see Appendix \ref{app:SNNSimul} for other parameters). Solid lines: theoretical $P_\nu(\omega)$ from Eq.~\eqref{eq:NuPSDunderAS} computed relying on the first 4096 modes of the FP operator spectral expansion. Shaded strips: mean $\pm \, 3$~SEM (standard error mean) of $P_\nu(\omega)$ from 10 simulations with different random synaptic matrices and same mean-field parameters. Power spectra are normalized by $N/\nu_0$, where $\nu_0$ is the average in time of $\nu_N(t)$ and $\nu_0 = \nu^*$ for theoretical $P_\nu(\omega)$. %
}
\label{fig:VIFNetTheoVsSimu}
\end{figure}

A comparison between Eqs.~\eqref{eq:LowDimERE} and \eqref{eq:DelayDistribERE} remarkably points out the equivalence of having, in a spiking neuron network, a non-instantaneous synaptic transmission or a suited distribution of axonal delays. In Eq.~\eqref{eq:DelayDistribERE} the expansion coefficients in $\vec{a}$ refer to the one-dimensional density $P(x,t)$, while in Eq.~\eqref{eq:LowDimERE} $\vec{a}_0$ is only an effective representation of the two-dimensional density $P(x,y,t)$. Nevertheless, provided that $\taus = \taud$, both the dynamics seen from the perspective of $\nu(t)$ are the same, as the functions $\Phi_0$, $\vec{f}_0$, $\vec{w}_0$, $\mathbf{W}_0$ and $\mathbf{\Lambda}_0$ are the same. In other words, under mean-field approximation and for not too large synaptic current fluctuations $\sigma_I$, having a local synaptic filtering with cut-off frequency $1/\tau_s$ is equivalent to have random axonal delays with exponential distribution \eqref{eq:ExpDelayDistrib} and decay constant $\taud = \taus$.

\section{Testing inertia equivalence in simulation}

The theory developed in the previous sections has been tested through extensive numerical simulations by using NEST \cite{Gewaltig2007} and the high-performance custom simulator implementing the event-based approach described in
\cite{Mattia2000}. The parameters for the numerical simulations have been identified following the procedure described in Appendix \ref{app:IdentifParam}. The network parameters corresponding to the results proposed in this section are summarized in Appendix \ref{app:SNNSimul}.

\subsection{Equivalence in the asynchronous state of finite-size networks}

A first evaluation of the effectiveness of the low-dimensional description derived above is possible by inspecting the activity in the presence of an endogenous noise in a network of a finite number $N$ of neurons trapped into a stable asynchronous state. 
As previously done in \cite{Mattia2002}, finite-size fluctuations can be incorporated into Eqs.~\eqref{eq:LowDimERE} and \eqref{eq:DelayDistribERE} as an additive forcing term to the expansion coefficient dynamics, giving rise to an equation for the emission rate $\nu_N(t)$ of a finite pool of neurons (see Appendix~\ref{app:thspectr} for details). The resulting dynamics can be linearized around the fixed-point $\nu^*$, allowing to compute the Fourier transform $\nu_N(\omega)$, and from it the power spectral density $P_\nu(\omega) = |\nu_N(\omega)|^2$, which turns out to be
 \begin{equation}
 P_\nu(\omega) = \frac{1 + 2 \, \mathrm{Re}\left[\vec{f}_0 \cdot (i \omega \mathbf{I} - \mathbf{\Lambda}_0)^{-1} \vec{\psi}_0\right]}%
 {\left|1 - \left[\Phi'_0 + i \omega \vec{f}_0 \cdot (i \omega \mathbf{I} - \mathbf{\Lambda}_0)^{-1} \vec{w}_0 \right] \rho(i \omega) \right|^2} \frac{\nu^*}{N} \, .
 \label{eq:NuPSDunderAS}
 \end{equation}
Here, $\mathbf{I}$ is the identity matrix, all the coefficients depending on $\nu(t)$ are now constants computed at $\nu(t)=\nu^*$, $\Phi'_0 = \partial_\nu \Phi_0$ and $\rho(i \omega)$ is the ratio between the Fourier transforms of $\mu_y$ and $\mu_I$, which results to be $\rho(i \omega) = 1/(1+i \omega \taus)$. Note that $\rho(i \omega)$ is the same as for the Fourier transform of the delay distribution \eqref{eq:DelayDistribERE} with $\taud = \taus$. If we consider an additional fixed transmission delay $\delta$, this transform generalizes to $\rho(i \omega) = \exp(-i \omega \delta)/(1+i \omega \taus)$ \cite{Mattia2002,Mattia2003}.

\begin{figure*}[!htb]
\includegraphics[width=1.0\textwidth]{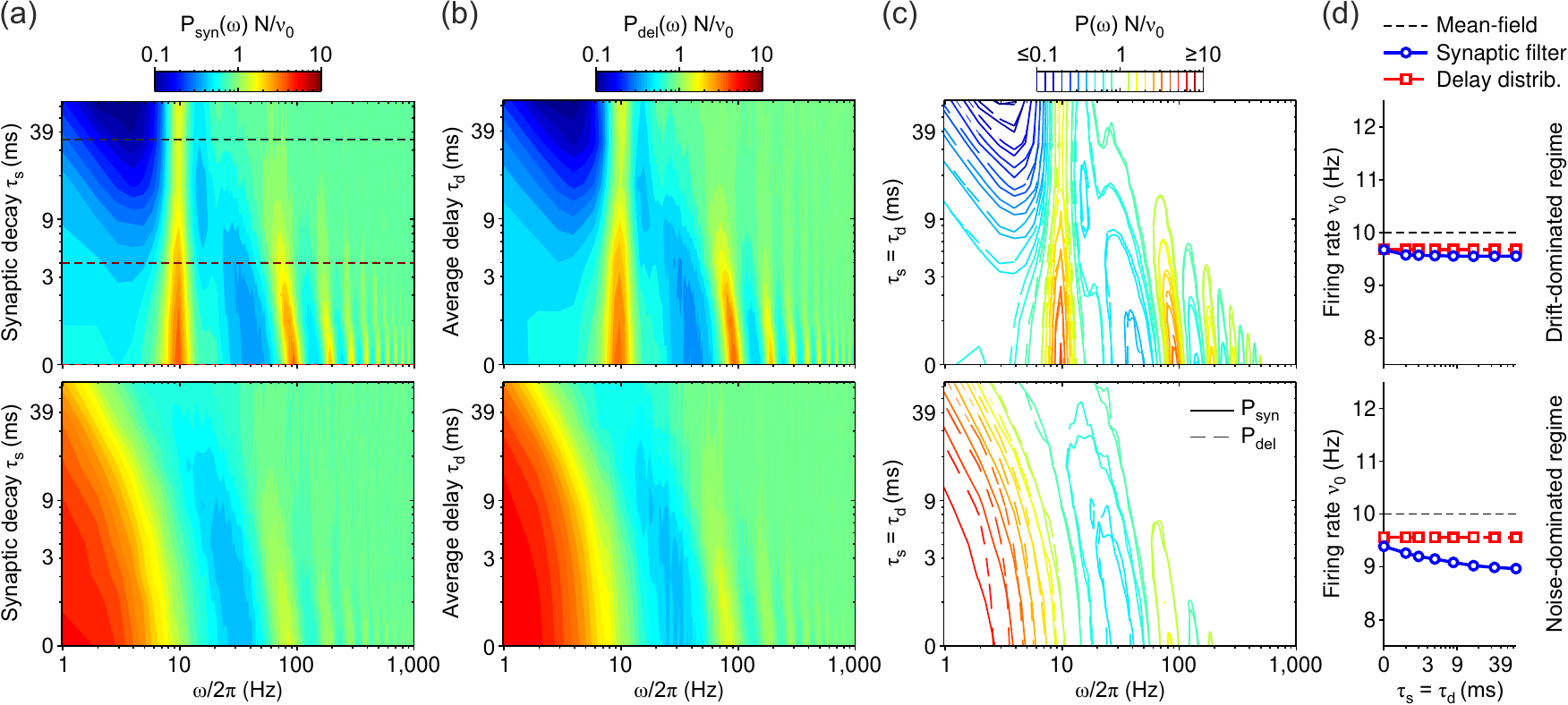} %
   \caption{ %
   Equivalence of non-instantaneous synaptic transmission (local inertia) and distribution of synaptic delays (distributed inertia) in a network of VIF neurons under drift- and noise-dominated regimes. %
(a) Power spectral density $P_\nu(\omega)$ of the simulated network activity in the case of non-instantaneous synaptic transmission, varying the synaptic time constant $\tau_{s}$. Dashed lines: $P_\nu(\omega)$ sections plotted in Fig.~\ref{fig:VIFNetTheoVsSimu}. %
(b) $P_\nu(\omega)$ for networks with an exponential distribution of spike transmission delays and with instantaneous synaptic transmission, varying the delay time constant $\tau_d$. In (a) and (b), top and bottom panels show the results for the network set in an asynchronous state in which neurons are in a drift- (top) and noise-dominated (bottom) regime, respectively. %
(c) Iso-power curves from panels (a) (solid lines) and (b) (dashed lines). %
(d) Average firing rate $\nu_0$ from the simulations used for the left panels with local (red plots) and distributed (blue plots) inertia. Error bars representing SEM are not visible. The not specified network parameters are as in Fig.~\ref{fig:VIFNetTheoVsSimu}. Also here, power spectra are averages across 10 simulations with different random synaptic matrices and same mean-field parameters.
} %
   \label{fig:VIFNetSynapticVsDelay}
 \end{figure*}

We compared this theoretical result with the $P_\nu(\omega)$ estimated 
from simulations of networks composed of simple IF neurons with synaptic filtering (see Appendix~\ref{app:C} for details), finding a remarkable agreement (Fig.~\ref{fig:VIFNetTheoVsSimu}). Here we resorted to the VIF neuron model \cite{Fusi1999}, an extended version of the widely used PIF neuron \cite{Gerstein1964}, for its amenability to analytical treatment \cite{Mattia2002} (see Appendix~\ref{app:SNNSimul} for details). %
Besides the good matching between theory and simulations, it is interesting to note how resonant peaks are differently affected by an increasing $\taus$. Not surprisingly, high-$\omega$ peaks are dampened due to the low-pass filtering of synaptic transmission. These resonances are related to the synaptic reverberation of spiking activity, which gives rise to the so-called \textit{transmission} poles in the linearized dynamics \cite{Treves1993a,Mattia2002}. In an excitatory network as in Fig.~\ref{fig:VIFNetTheoVsSimu}, these are expected to be found at frequencies multiples of the inverse of the average time needed by a presynaptic spike to affect postsynaptic potential, here roughly the same as $1/(\delta + \taus)$. On the other hand, the low-$\omega$ peaks do not display any shift in frequency, although their power is dampened as well. This is because such peaks are due to the so-called \textit{diffusion} poles in the linearized dynamics \cite{Spiridon1999,Mattia2002}. They occur when neurons emit spikes at drift-dominated (suprathreshold) regime. In this regime, the distribution of the inter-spike intervals is narrow and resonant peaks emerge at $\omega/2\pi$ multiples of $\nu^*$ (equal to 10~Hz in Fig.~\ref{fig:VIFNetTheoVsSimu}). We remark that, due to the shifting at low-$\omega$ of the transmission peaks, a constructive interference with diffusion peaks may occur. This explains the non-monotonic change of power of the second diffusion peak at 20~Hz in Fig.~\ref{fig:VIFNetTheoVsSimu}. Indeed, its power decreases when $\taus$ is increased from 0~ms to 4~ms, as expected, whereas due to the mentioned interference it is heightened for $\taus = 32$~ms.  This is because the lowest transmission peak in this case is expected to be found at $\omega/2\pi = 23.8$~Hz, not too far from $2 \, \nu^* = 20$~Hz.

\subsection{Equivalence under noise- and drift-dominated regime}
\label{sec:EquivNDRegimes}

To further test the equivalence between local and distributed inertia, we investigated whether this match worked well not only for networks of neurons under drift-dominated spiking regime, as shown in Fig.~\ref{fig:VIFNetTheoVsSimu}, but also under fluctuation-driven regimes. Indeed, synaptic filtering may have significantly different effects in these two regimes \cite{Brunel2001,MorenoBote2004}.

\InRefFig{fig:VIFNetSynapticVsDelay} shows for comparison the power spectral densities $P_\nu(\omega)$ obtained by varying $\taus$ and $\taud$ in the same excitatory VIF neuron networks in which either non-instantaneous synaptic transmission or a distribution of transmission delays was incorporated (Fig.~\ref{fig:VIFNetSynapticVsDelay}a and b, respectively). Networks with neurons spiking at drift-dominated regimes (top panels) were the same as those shown in Fig.~\ref{fig:VIFNetTheoVsSimu}. A remarkable agreement between simulations with local and distributed inertia was confirmed as iso-power curves tightly overlapped (see Fig.~\ref{fig:VIFNetSynapticVsDelay}c) for a wide range of filter timescales. The same good match between these two network types was apparent also under noise-dominated regime (bottom panels).

\begin{figure*}[!htb]
\includegraphics[width=1.0\textwidth]{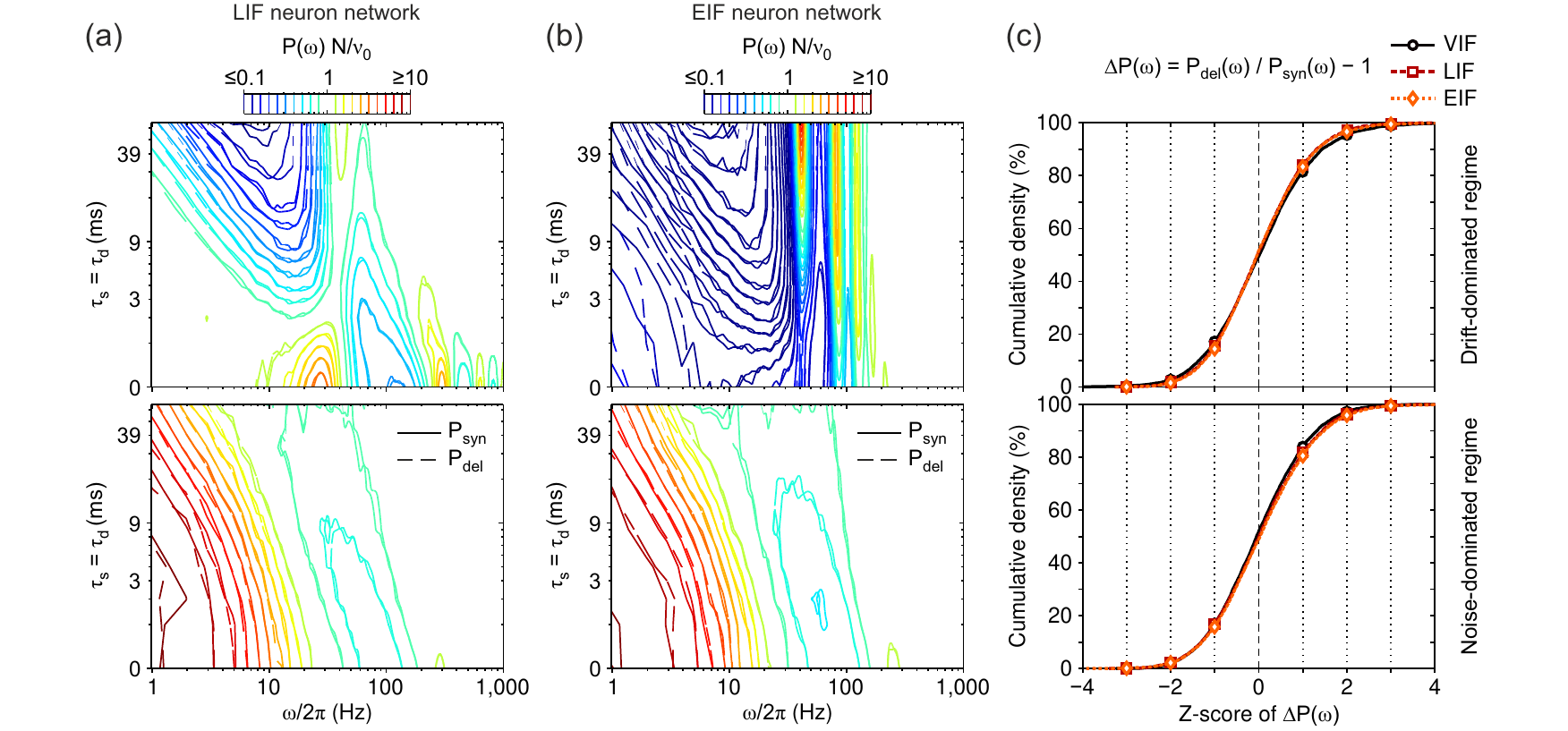} %
\caption{ %
Match between relative power spectral densities $P_\nu(\omega) \, N /\nu_0$ in networks of leaky (a) and exponential (b) integrate-and-fire excitatory neurons (LIF and EIF, respectively) with non-instantaneous synaptic transmission (solid lines) and distributions of synaptic delays (dashed lines).
Network parameters are set in order to have an asynchronous state with average firing rate $\nu_0 = 40 \, \mathrm{Hz}$. 
Contour lines as in \RefFig{fig:VIFNetSynapticVsDelay}c. 
$P_\nu(\omega)$ are averages across 10 simulations with different random synaptic matrices and same mean-field parameters.
(c) Cumulative distribution of relative differences $\Delta P(\omega) = P_\mathrm{del}(\omega)/P_\mathrm{syn}(\omega) - 1$ across different time scales ($\taus$ and $\taud$) and Fourier frequencies $\omega$ from panels (a) and (b) for LIF (red) and EIF (orange) neuron networks, respectively. As a reference, also the same cumulative distribution for VIF neuron networks (obtained from $P_\nu(\omega)$ in \RefFig{fig:VIFNetSynapticVsDelay}c) is plotted (black). $\Delta P(\omega)$ are measured as $z$-scores, i.e. averages divided by standard deviations of values across simulations.
Top and bottom panels: networks working under drift-dominated and noise-dominated regime, respectively (see Appendix \ref{app:SNNSimul}). %
}
\label{fig:LIFEIF}
\end{figure*}

As expected in this regime, the diffusion poles become real numbers \cite{Mattia2002} and resonant peaks at frequency $\omega/2\pi$ multiples of firing rate $\nu^*$ disappear, although the high-$\omega$ resonances corresponding to transmission poles remain almost unaffected by this change of regime.
It interesting to point out here that widening the filtering time window (i.e., increasing $\taus$ and $\taud$) leads to almost flat power spectra, compatibly with the flattening of the response amplitude of isolated neurons receiving colored noise as input currents \cite{Brunel2001,Schuecker2015}.

We remark that in Fig.~\ref{fig:VIFNetSynapticVsDelay}a-c comparisons rely on normalized spectra, i.e. $P_\nu(\omega) \, N/\nu_0$. On the one hand, this means that power spectra shapes does not depend on the particular inertia implemented in the network. On the other hand, this does not guarantee that average firing rates $\nu_0$ are the same and do not change with the inertia-related timescales. To address this issue, in Fig.~\ref{fig:VIFNetSynapticVsDelay}d the measured $\nu_0$ are compared. As expected, the average firing rate when distribution of delay are incorporated does not change by varying $\taud$. Mean-field fixed-point $\nu^*$ mildly overestimates $\nu_0$ and this is due to the assumptions underlying the diffusion approximation. Indeed, the difference $|\nu_0 - \nu^*|$ shrinks as the number of synaptic contacts per neuron and the network size $N$ increase by keeping unchanged drift and diffusion coefficients of the related Langevin equation (this result is not shown in the figures). But what is important to note here is rather that the average firing rate $\nu_0$ measured for local and distributed inertia are different, and the difference increases with $\taus=\taud$. This trend is particularly apparent under noise-dominated regime, and it is not completely unexpected. Indeed, in the small-$\taus$ limit \cite{Brunel1998,MorenoBote2004,Schuecker2015} and in the long synaptic timescale condition \cite{MorenoBote2004} macroscopic differences are known to exist. This phenomenon is not captured by our 0-th order perturbative approach, since the total flux of realizations crossing the emission threshold $\vthr$, which contributes to the total emission rate, depends on the shape of the probability current distribution $\mathcal{F}_x(\vthr,y) = \partial_x P|_{x= \vthr} = \nu(y,t)$ 
\cite{Brunel1998,Schuecker2015}. 
Indeed, as shown in \InRefFig{fig:TimeEvolutionOfVandI}c-d, the section of $P(x,y,t)$ close to emission threshold $\vthr$ shrinks when $\taus$ increases from 4~ms to 64~ms.%
In our perturbative approach this dependence is completely neglected, as the distribution is assumed to be a Dirac's $\delta$.

\subsection{Independence from spiking neuron models}

The developed theory is of general applicability, i.e. it applies to a wide range of spiking neuron models. As a result, the equivalence between local and distributed inertia proved for a network of simplified VIF neurons together with the theoretical expression for the power spectra of $\nu(t)$ are both expected to hold also for networks of more realistic single-cell models like the leaky IF (LIF, \cite{Tuckwell1988}) and the exponential IF (EIF, \cite{Fourcaud-Trocme2003}) neurons.

To verify this expectation, we directly simulated networks of LIF and EIF neurons with a mean-field stable equilibrium point at $\nu^*= 40 \, \mathrm{Hz}$. In \RefFig{fig:LIFEIF} the mean $P_\nu(\omega) \, N / \nu_0$ for these kinds of networks is compared both under drift-dominated (top panels) and noise-dominated (bottom) regimes. A remarkable agreement is apparent for both LIF and EIF neuron networks (\RefFig{fig:LIFEIF}a and b, respectively), confirming the generality of the developed approach. To further test the absence of any bias due to the specific neuron model chosen, we computed the relative differences between the power spectral densities when non-instantaneous synaptic filtering ($P_\mathrm{syn}$) and transmission delay distribution ($P_\mathrm{del}$) are taken into account. The cumulative distributions of these discrepancies across all tested inertia time scales ($\taus = \taud$) and Fourier frequencies ($\omega/2\pi$) are shown in \RefFig{fig:LIFEIF}c. Interestingly, no significant differences are visible for the three chosen models (VIF, LIF, EIF), further proving that in all regimes (drift- and noise-dominated) our dimensional reduction does not depend on the specific single neuron dynamics.

Starting from this, it is not surprising to note here that spectral features similar to those highlighted in \RefFig{fig:VIFNetSynapticVsDelay} for VIF neuron networks are also displayed by LIF and EIF neuron networks. More specifically, resonant peaks at multiple frequencies of $\nu_0$ under drift-dominated regimes and resonances at higher-$\omega$ due to the transmission poles -- i.e., those related to the average spike transmission delay -- are also visible in both \RefFig{fig:LIFEIF}a and b in LIF and EIF neuron networks, respectively.

\subsection{Equivalence beyond the asynchronous state}

So far, we tested the validity of the equivalence between non-instantaneous synaptic transmission and delay inhomogeneity in linearizable dynamical regimes like the asynchronous state. To further assess the generality of our theoretical results, we simulated an inhibitory network of LIF neurons (see Appendix \ref{app:SNNSimul}) undergoing a transition from a point-attractor dynamics to a regime of collective oscillations through a supercritical Hopf bifurcation as in \cite{Brunel1999}.

\begin{figure}[!ht]
\includegraphics[width=1.0\columnwidth]{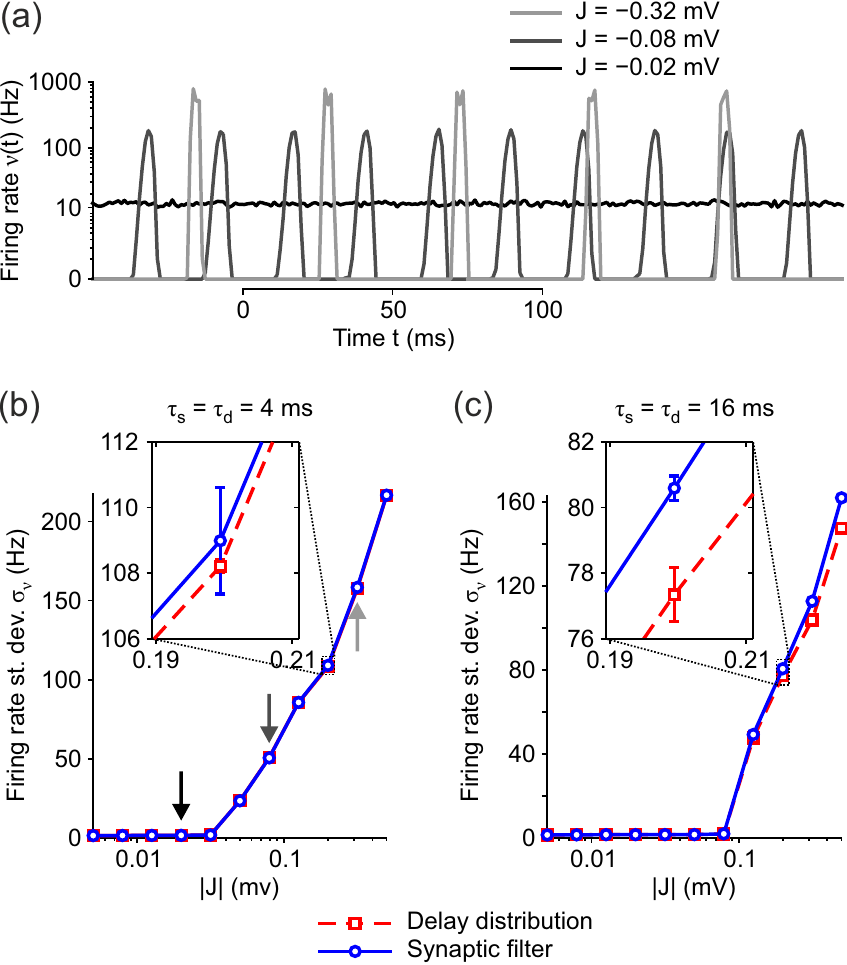} %
\caption{ %
Comparison between emission rate dynamics in networks of inhibitory LIF neurons with both synaptic filtering and transmission delay distribution outside the asynchronous state. Networks undergo a supercritical Hopf bifurcation by progressively increasing the synaptic strength $|J|$. %
(a) Transition from a stable asynchronous state (black line, $J = -0.02 \, \mathrm{mV}$) to a global oscillation regime (light and dark gray lines, $|J| > 0.04 \, \mathrm{mV}$). 
Stable and unstable equilibrium points are always set at $\nu = 10 \, \mathrm{Hz}$ by compensating the increase in synaptic inhibition with a growing external excitatory current $I_\mathrm{ext}$. Synaptic filtering with $\taus = 4 \, \mathrm{ms}$ is incorporated. %
(b-c) Two inertia time scales have been tested ($\taud = \taus = 4 \, \mathrm{ms}$ (b) and $16 \, \mathrm{ms}$ (c)) by comparing the standard deviation $\sigma_\nu$ of the instantaneous firing rate $\nu$ across the whole simulation time of 60 s (the first second of the time series is discarded to take into account only the asymptotic dynamics).
Blue solid and red dashed lines, $\sigma_\nu$ averaged across 10 mean-field equivalent simulations with random synaptic matrices and incorporating equivalent synaptic filtering and transmission delay distribution, respectively. Error bars, standard error mean.
Shaded arrows in (b) mark the examples in panel (a).
Both $J$ and $I_\mathrm{ext}$ are the same in panels (b) and (c) and their values are detailed in Appendix~\ref{app:SNNSimul}, together with the other network parameters.
}
\label{fig:Hopf}
\end{figure}

\InRefFig{fig:Hopf}a shows three examples of population activity $\nu(t)$ for increasing magnitude $|J|$ of the inhibitory coupling when a non-instantaneous synaptic transmission with decay time $\taus = 4 \, \mathrm{ms}$ is incorporated. For relatively weak couplings (lowest $|J|$), the network is in an asynchronous state and its activity fluctuates around the equilibrium point $\nu^*=10$~Hz (black line). By increasing the synaptic inhibition above a threshold value ($|J| > 0.04$~mV), the network starts to display coherent global oscillations of the emission rate with increasing amplitude (light and dark gray lines). 

We spanned the same range of synaptic strength $J$ making the Hopf bifurcation apparent in networks with both synaptic filtering and delay distribution, having care to use the same mean-field parameters (\RefFig{fig:Hopf}b). For each $J$ we carried out the related external excitatory current $I_{ext}$ needed to keep unchanged the fixed-point at $\nu^*=10$~Hz regardless of its stability (see details in Appendix \ref{app:SNNSimul}). We compared the dynamics of the two network types by measuring the standard deviation $\sigma_\nu$ across time of the emission rate $\nu(t)$ under stationary conditions (i.e., discarding the first second of the time series). As a remarkable result, we found an almost perfect match between $\sigma_\nu$ measured in networks with both local and distributed inertia. Indeed, the relative mismatch between $\sigma_\nu$ is always lower than $1\%$ (compare solid and dashed lines). This also implies that regardless of the inertia type, the transitions across regimes roughly occurs at the same critical value ($J \simeq - 0.04$~mV) with a discontinuity of the derivative $\partial \sigma_\nu / \partial J$, highlighting in both cases the same supercritical Hopf bifurcation. Similar results were obtained for longer filtering timescales ($\taus = \taud = 16$~ms, \RefFig{fig:Hopf}c), although in this case significant $\sigma_\nu$ differences can be measured for large enough $J$.

Altogether, these results further prove the generality of the theoretical equivalence derived in Sec. \ref{sec:localvsdistrib}, clearly pointing out that local and distributed inertia give rise to the same collective dynamics not only in a tameable condition like the linearizable asynchronous state, but also for nonstationary network states like the global oscillatory regime.

\section{Discussion}

Local and distributed inertia in networks of spiking neurons are equivalent,  provided that non-instantaneous post-synaptic currents from incoming spikes (\textit{local inertia}) have the same time course as the distribution of spike transmission delays across axons and dendrites (\textit{distributed inertia}). This is one of the main findings of this work, and we proved such equivalence by developing a novel population density approach to the network dynamics of neurons with non-Markovian membrane potentials. This general method consists of a dimensional reduction of the two-dimensional population density dynamics arising from the Markovian embedding \cite{Risken1984,Hanggi1985} of the membrane potential evolution. The effective one-dimensional description we obtained relies on both an extended mean-field approximation \cite{Amit1997} and the assumption of a relatively low synaptic noise, without requiring any additional constraint on the correlation time due to the incorporated non-instantaneous synaptic transmission. This is our second main result, from which we obtained the same emission rate equation (ERE) as the one arising from the spectral expansion of the one-dimensional FP equation for the population density in the absence of synaptic filtering \cite{Mattia2002}, but with an important difference; drift and diffusion coefficients, related to mean and variance of the input currents, respectively, now depend on a low-pass filtered version of the instantaneous firing rate of the network. We tested both the generality of the developed theory and the equivalence between local and distributed inertia through extensive numerical simulations, finding a remarkable agreement with theoretical expectations in a variety of dynamical scenarios, including noise- and drift-dominated regimes, equilibrium states and collective oscillations, excitatory and inhibitory synaptic connections, and for a wide set of spiking neuron models.

\subsection{Comparison with previous studies}

Reducing into an effective one-dimensional Markovian problem the dynamics of a non-Markovian system -- which in turn can be embedded into a larger set of coupled Markovian processes --  has a long history in statistical physics \cite{Risken1984,Hanggi1995}. Of course, the resulting approximation schemes have unavoidable limitations in representing the true dynamics of the original non-Markovian problem. Nevertheless, many successes were accumulated in the late 80s of the past century \cite{Sancho1982,Hanggi1985,Fox1986,Grigolini1986,Jung1987,Masoliver1987a}. More specifically, many of them focused on the nonstationary dynamics underlying the specific problem of the first-passage time (FPT), thus taking into account an absorbing barrier, similarly to the spike emission threshold in neuronal modeling. Remarkably, also in this FPT framework stochastic processes driven by `colored noise' can be effectively approximated by one-dimensional Markovian processes, provided that the related FP equations incorporate a state-dependent nonlinear diffusion coefficient \cite{Hanggi1985,Masoliver1987a}. This coefficient is not time-varying only in the limit of small noise and/or small correlation times $\taus$ \cite{Fox1986,Grigolini1986}. Other perturbative approaches have been worked out to cover the long-$\taus$ range \cite{Jung1987,Hwalisz1989}, but the effectiveness of such Markovian approximations is mainly limited by the lack of a proper management of the boundary conditions \cite{Doering1987,Hanggi1995,Klosek1998}. Intriguingly, here by including the proper absorbing barrier prescription due to the presence of a forcing white noise with arbitrarily small size $\sigma_{ext}$ (similarly to \cite{Moreno2002}), the theoretical derivation of the approximated ERE \eqref{eq:LowDimERE} eventually recovers a representation with time-varying state-dependent diffusion coefficients (see for instance Section V.B of \cite{Hanggi1995}). Indeed, all the coefficients of this ERE depend on a filtered version $\mu_y(t)$ of the total emission rate $\nu(t)$, which from Eq.~\eqref{eq:TotalEmissionRate} is in turn a nonlinear functional of the probability density $P(x,y,t)$.

The approximated population density approach here developed brought also another interesting result. Typically, power spectra and Fourier transfer functions of the population activity in network of spiking neurons and synaptic filtering are characterized outside biologically relevant regimes. This is because theoretical approaches are perturbative and target extreme conditions, in which either fast \cite{Brunel2001,Fourcaud2002,Schuecker2015} or slow \cite{MorenoBote2004,Moreno-Bote2006a} synapses are considered, or in which neurons work under strong drift-dominated (low-noise) regimes \cite{Lindner2004b,Schwalger2015}. Here, we bridge this gap by presenting a theoretical description valid for both noise- and drift-dominated regimes, and for the whole spectrum of synaptic time scales and Fourier frequencies (\RefFigs{fig:VIFNetTheoVsSimu}--\ref{fig:LIFEIF}).

It is also interesting to remark that the proved equivalence between local and distributed inertia allows to map a many-body system with non-Markovian coupled elements onto another many-body system where many Markovian units have delayed interactions. In the latter framework, different types of collective dynamics and, in particular, synchronization have been widely studied \cite{Boccaletti2006}. The related theoretical results are then expected to be exportable to the case in which an equivalent synaptic filtering is taken into account. 
An intriguing example is the use of effective time-delayed feedback loops to control neuronal pathological states like the synchronous firing observed in Parkinson's disease or epilepsy \cite{Schiff1994,Rosenblum2004}.
The same strategy could be implemented by relying on a suitably slow synaptic feedback, by taking advantage of the natural plasticity of the inter-neuronal couplings. Due to this neurophysiological adaptability, the optimal delayed feedback needed to control pathological dynamics could be in principle `learned', similarly to what occurs in `echo-state networks' and `liquid-state machines' \cite{Jaeger2004,Maass2007,Sussillo2009a}.

\subsection{Dimensional reduction}

Other dimensional reductions of the Fokker-Planck dynamics arising when synaptic filtering is incorporated in network of spiking neurons have been proposed in the past to obtain effective kinetic representations amenable to numerical integration \cite{Cai2004,Rangan2006,Ly2007}. They mainly rely on the centered moment closure method usually applied to kinetic problems in statistical physics \cite{Levermore1996}. Drawbacks of this approach are related to their limited applicability to noise-dominated regimes \cite{Ly2007,Ly2013}, which are of particular relevance in neuroscience, being associated to activity states with balanced excitation and inhibition \cite{VanVreeswijk1996,Shu2003,Yizhar2011}. Even when some of these limitations are removed by introducing modifications to the standard mean-field approximation \cite{Ly2013}, this kind of dimensional reduction mainly remains a computational method in which, in addition to the dynamics of the centered moments, a one-dimensional partial differential equation has to be numerically integrated. 
On the contrary, our approach, in which a spectral expansion of the two-dimensional FP operator is operated together with a suited centered moment closure, is not only amenable to numerical integration. For instance, theoretical insights on the dynamics of the investigated system can be effectively carried out \cite{Mattia2002,Mattia2004,Schaffer2013} by focusing on the slowest modes of the expansion.

\subsection{Limitations}

The developed theoretical description of the network dynamics with synaptic currents is approximated, and as such it is subject to several limitations. Our starting point is a population density approach which, like others \cite{Knight1996,Brunel1999,Mattia2002}, relies on both a diffusion and a mean-field approximation. The underlying hypotheses require that each neuron receives a large amount of spikes per unit time, and that postsynaptic currents due to single spikes induce small changes of the membrane potential. These conditions are well satisfied in neocortex \cite{Tuckwell1988,Amit1997}.

To reduce the dimensionality of the population density dynamics, we further resorted to two simplifications. Firstly, to safely manage the boundary conditions in the spectral expansion, we introduced a forcing white noise current with an arbitrarily small size $\sigma_{ext}$ (as in \cite{Moreno2002}). The intrinsic stochasticity of ionic channels of neuronal membrane potential can be the source of such additional input current \cite{Linaro2011,Goldwyn2011}. But, even in the limit $\sigma_{ext} \to 0$, we expect to find a good agreement between our theory and simulations. Indeed, although $\sigma_{ext}$ widely varied in the various comparisons presented in \RefFigs{fig:VIFNetTheoVsSimu}--\ref{fig:Hopf}, the agreement between simulations with synaptic filtering and delay distribution did not show any significant difference. However, it is important to remark that, in the absence of this external noise, the spectral expansion we used is no more valid, as it relies on the eigenfunctions of a FP operator with $P(\vthr,y,t) = 0$. This boundary condition is not valid if $\sigma_{ext} = 0$ \cite{Doering1987,Klosek1998,Brunel1998,Schuecker2015}, in which case a different basis for the probability density decomposition has to be adopted.  

The second strong simplification we implemented was to consider small enough the fluctuation size $\sigma_I$ of the synaptic currents. Such a low-noise regime gave us the possibility to assume the marginal distribution of the current $I(t)$ narrowly distributed around its time-varying mean $\mu_y(t)$. As a result, the $V-I$ dynamics were reduced to a one-dimensional FP equation centered around $I(t) = \mu_y(t)$. Rather surprisingly, such a rough approximation for which the shape of the marginal distribution $\rho_y(y,t)$ is completely neglected, seems to work remarkably well. This is a result that we suspect being due to the symmetry of the narrow distribution $\rho_y(y,t)$, such that contributions of the terms with $I > \mu_y$ are almost perfectly compensated by those with $I < \mu_y$. The only main discrepancy we found is in the mean firing rate, which our approximated theory systematically overestimated (see \RefFig{fig:VIFNetSynapticVsDelay}d). This is an expected error, since synaptic filtering is known to mildly reduce the emission rate of spikes in both the small- \cite{Brunel1998,MorenoBote2004,Schuecker2015} and the large-$\taus$ \cite{MorenoBote2004} limit. Such discrepancy should disappear in our approximated development if higher-orders of the centered moment closure were incorporated. Indeed, this would allow to take into account also other features of $\rho_y(y,t)$, such as its standard deviation and skewness.

\begin{acknowledgments}
Work partially supported by the University of Genoa (M.S.) and by the EU Horizon 2020 Research and Innovation Programme under HBP SGA1 (grant no. 720270 to M.M.).
\end{acknowledgments}

%

\section{Appendix}

\subsection{Derivation of the power spectral density $P_\nu(\omega)$ of $\nu_N(t)$}
\label{app:thspectr}

In a network of a finite number $N$ of neurons, the instantaneous firing rate $\nu_N(t)$ display endogenous fluctuations. Indeed, due to the finite number of spikes emitted by such network in relatively small time intervals, the counting Poissonian statistics makes $\nu_N(t)$ well described by a stochastic variable whose variance scales as $1/N$ \cite{Spiridon1999,Brunel1999,Mattia2002}. These activity fluctuations can be seen as an ongoing stimulation of the infinite-size network and can be obtained by introducing an effective stochastic driving force different for each eigenmode of the FP operator \cite{Mattia2002}. Following this approach and generalizing Eq.~\eqref{eq:LowDimERE} to the case of a finite-size network, we obtain the following stochastic emission-rate equations
\begin{equation}
 \begin{cases}
 \dot{\vec{a}}_0 = \left(\mathbf{\Lambda}_{0} + \mathbf{W}_{0} \, \dot{\mu}_y \right) \, \vec{a}_0 + \vec{w}_0 \, \dot{\mu}_y + \vec{\psi}_0 \, \eta_N\\
 \dot{\mu}_y = \displaystyle \left(\mu_I(\nu_N) - \mu_y\right)/\tau_s \\
 \nu = \Phi_{0} + \vec{f}_0 \cdot \vec{a}_0 \\
 \nu_N = \nu + \eta_N
 \end{cases} \, ,
 \label{eq:FiniteSizeLowDimERE}
 \end{equation}
 where $\eta_N(t)$ models the finite-size fluctuations of the instantaneous firing rate $\nu(t)$ in the infinite-size limit. For large enough $N$, $\eta_N$ is well approximated by a Gaussian memoryless white noise, with zero mean and variance $\nu(t)/N$. In the above equations the coefficients depend on $\nu_N(t)$ instead of $\nu(t)$, as the infinitesimal mean $\mu_I$ and variance $\sigma^2_I$ are functions of the instantaneous spike rate of the presynaptic neurons. The additional coefficients $\vec{\psi_0} = \{\psi_{n}(\vres,y)|_{y=\mu_y}\}$ result from having incorporated finite-size fluctuations to the boundary condition on the flux conservation of the realizations $V(t)$ exiting from $\vthr$ and reentering in $\vres$ \cite{Mattia2002}.
 
In the $N\to\infty$ limit, the network dynamics can be set into an asynchronous state such that $\nu(t) = \nu^*$ is a fixed-point [$\nu^* = \Phi_0(\nu^*)$] with local stability [$\Phi_0'(\nu^*)<1$] and the single-neuron spiking activity is asynchronous \cite{Amit1997,Brunel2000}. In this state, finite-size fluctuations bring $\nu_N(t)$ to wander around the fixed-point $\nu^*$, and Eq.~\eqref{eq:FiniteSizeLowDimERE} can be linearized by neglecting the terms of order higher than $\mathcal{O}(\eta_N)$ in the Taylor's series expansion around $\nu(t) = \nu^*$. From this linearized dynamics, the Fourier transform $\nu_N(\omega)$ can be obtained (see \cite{Mattia2002,Mattia2004} for details), and the power spectral density $P_\nu(\omega) = |\nu_N(\omega)|^2$ turns out to be
 \begin{equation}
 P_\nu(\omega) = \frac{1 + 2 \, \mathrm{Re}\left[\vec{f}_0 \cdot (i \omega \mathbf{I} - \mathbf{\Lambda}_0)^{-1} \vec{\psi}_0\right]}%
 {\left|1 - \left[\Phi'_0 + i \omega \vec{f}_0 \cdot (i \omega \mathbf{I} - \mathbf{\Lambda}_0)^{-1} \vec{w}_0 \right] \rho(i \omega) \right|^2} \frac{\nu^*}{N} \, ,
 \end{equation}
which is the same expression detailed in Eq.~\eqref{eq:NuPSDunderAS}.

\subsection{Identification of mean-field parameters}
\label{app:IdentifParam}

To find the parameters for the numerical simulations (with NEST
\cite{Gewaltig2007} and the high-performance custom simulator
implementing the event-based approach described in
\cite{Mattia2000}), the following procedure has been adopted.

An emission rate fixed point $\nu^{*}$ was fixed a priori, and
the contour line in the $(\mu,\sigma^{2})$ plane defined by
$\Phi(\mu,\sigma^{2}) = \nu^{*}$ was numerically determined.
$\Phi(\mu,\sigma^{2})$ was computed analytically for the VIF
\cite{Fusi1999} and LIF \cite{Ricciardi1979,Amit1997}
neuron models. For the EIF
neuron model, in the absence of an analytically expression for
$\Phi(\mu,\sigma^{2})$ valid in all the considered conditions,
we used a numerical cubic interpolation (using the Matlab 
-- The MathWorks, Natick, MA --
function \texttt{interp2}) passing through samples obtained
from NEST simulation data. Depending on the regime of interest (noise- or
drift-dominated), a proper point along this 
iso-frequency
line was chosen.

Once determined $\mu$ and
$\sigma^{2}$, also $J$ can be determined by imposing the
value of $\frac{\mathrm{d}}{\mathrm{d}\nu}
\Phi(\mu,\sigma^{2})$ at the fixed point $\nu^{*}$, 
which is directly related to its degree of stability \cite{Mattia2002}. 
Indeed, 
by taking into account \RefEq{eq:moments},
it is sufficient to solve the
following second-order equation in $J$:
\begin{equation}
    \frac{\mathrm{d}}{\mathrm{d}\nu} \Phi(\mu,\sigma^{2}) = C J \left(\frac{\partial\Phi}{\partial\mu} + J\frac{\partial\Phi}{\partial\sigma^{2}}\right)
\end{equation}
\noindent where both $\partial\Phi/\partial\mu$ and
$\partial\Phi/\partial\sigma^{2}$ can be suitably computed from
$\Phi(\mu,\sigma)$. Of the possible solutions, we take the
one corresponding to the stable fixed point $\Phi(\mu,\sigma^{2}) = \nu^*$ with the highest firing rate, since it is related to the proper range
of $\nu$ values.

The final step is to determine mean $\mu_{ext}$ and variance $\sigma^2_{ext}$ of the external current $I_{ext}$ to be added to the recurrent synaptic contribution in order to obtain the chosen $\mu$ and $\sigma^2$. 
$I_{ext}$ is the sum of a constant current bias $I_{DC}$ and a Poissonian spike train from $C_{ext}$ independent sources firing at rate $\nu_{ext}$ through an instantaneous synaptic transmission with efficacy $J_{ext}$. Under diffusion approximation, $I_{ext}$ is a memoryless Wiener process and by imposing the spike rate $C_{ext} \, \nu_{ext}$ from the external neurons, the two remaining parameters $I_{DC}$ and $J_{ext}$ are uniquely determined, in the distribution of delays case, by solving 
the following system
\begin{equation}
\label{e:mean_var_stat}
    \begin{cases}
    \mu = J C \nu^{*} + J_{ext} C_{ext} \nu_{ext} + I_{DC}\\
    \sigma^{2} = J^{2} C \nu^{*} + J^{2}_{ext} C_{ext} \nu_{ext}
    \end{cases} \, .
\end{equation}
When the non-instantaneous synaptic transmission is incorporated, the recurrent synaptic efficacy has to be simply rescaled by the time constant $\taus$, i.e. $J \to J/\taus$. 
This is the approach used to design all the VIF, LIF and EIF neuron networks analyzed in Figs. \ref{fig:VIFNetTheoVsSimu}-\ref{fig:LIFEIF}.

For the results shown in \RefFig{fig:Hopf}, we used the
following protocol. We started with the simulation of a LIF inhibitory neuron network in an asynchronous state, thus determining $J$, $I_{DC}$ and $J_{ext}$ as stated above. Then, we increased $|J|$ and kept unchanged the (stable or unstable) fixed-point $\nu = \nu^*$ by increasing $I_{DC}$ accordingly.

\subsection{Spiking neuron network simulations}
\label{app:SNNSimul}

All the simulated IF neuron networks under asynchronous state (Figs.~\ref{fig:VIFNetTheoVsSimu}-\ref{fig:LIFEIF}) are modeled by considering a sparse recurrent connectivity with connection probability $C/N$, being $C$ the average number of synapses per neuron. Each neuron receives an external input from independent Poissonian spike trains with frequency $C_{ext} \, \nu_{ext}$ through instantaneous synapses with  efficacy $J_{ext}$. Spikes are always delivered to post-synaptic neurons with a minimum delay $\delta_{min}$. The values for both the synaptic time constant $\taus$ and the decay constant $\taud$ of the exponential delay distribution are ${0,1,2,4,8,16,32,64}$~ms.

\begin{table}[!ht]
\centering
\begin{tabular}{|m{6cm}|m{1cm}|m{1cm}|}
\hline
\textbf{Parameter} & \textbf{Value (DD)} & \textbf{Value (ND)} \\
\hline Number of excitatory neurons $N$ & 2000 & 2000 \\ 
\hline
Mean excitatory synapses per neuron $C$ & 100 & 100 \\ 
\hline 
Excitatory PSP amplitude due to recurrent spikes $J$ [$\vthr$] & 0.0075 & 0.014 \\
\hline 
Minimum transmission delay $\delta_{min}$ [ms] & 10 & 10 \\
\hline
External spike rate $C_{ext} \, \nu_{ext}$ [Hz] & 3000 & 3000 \\ 
\hline
Excitatory PSP amplitude due to external spikes $J_{ext}$ [$\vthr$] & 0.0114 & 0.0726 \\ 
\hline
Current bias $I_{DC}$ [$\vthr/$ms] & $-0.032$ & $-0.242$ \\ 
\hline
Refractory period $\tau_{ref}$ [ms] & 0 & 0 \\ 
\hline
Firing threshold $\vthr$  & 1 & 1 \\
\hline
Resting potential $V_E$ [$\vthr$] & 0 & 0 \\ 
\hline
Reset potential $\vres$ [$\vthr$] & 0 & 0 \\ 
\hline
Firing rate $\nu*$ [Hz] & 10 & 10 \\
\hline
\end{tabular}
\caption{Parameters of the VIF neuron networks at
drift-dominated (DD) and noise-dominated (ND) regimes.}
\label{tab:VIF}
\end{table}

\begin{table}[!ht]
\centering
\begin{tabular}{|m{6cm}|m{1cm}|m{1cm}|}
\hline
\textbf{Parameter} & \textbf{Value (DD)} & \textbf{Value (ND)} \\
\hline Number of excitatory neurons $N$ & 2000 & 2000 \\ 
\hline
Mean excitatory synapses per neuron $C$ & 100 & 100 \\ 
\hline Excitatory PSP amplitude due to recurrent spikes $J$ [mV] & 0.140 & 0.213 \\
\hline 
Minimum transmission delay $\delta_{min}$ [ms] & 3 & 3 \\
\hline
External spike rate $C_{ext} \, \nu_{ext}$ [Hz] & 12000 & 12000 \\ 
\hline
Excitatory PSP amplitude due to external spikes $J_{ext}$ [mV] & 0.272 & 1.518 \\ 
\hline
Current bias $I_{DC}$ [nA] & $-1.23$ & $-9.16$ \\ 
\hline
Membrane time constant $\tau_m$ [ms] & 20 & 20 \\ 
\hline
Refractory period $\tau_{ref}$ [ms] & 0 & 0 \\ 
\hline
Firing threshold $\vthr$ [mV] & 20 & 20 \\
\hline
Membrane capacitance $C_m$ [pF] & 500 & 500 \\ 
\hline
Resting potential $V_E$ [mV] & 0 & 0 \\ 
\hline
Reset potential $\vres$ [mV] & 0 & 0 \\ 
\hline
Firing rate $\nu*$ [Hz] & 40 & 40 \\
\hline
\end{tabular}
\caption{Parameters of the simulated LIF neuron networks at
drift-dominated (DD) and noise-dominated (ND) regimes.}
\label{tab:LIF}
\end{table}

The VIF neuron networks are composed of the `VLSI' integrate-and-fire neurons introduced in \cite{Fusi1999}. This model neuron is an extended version of the standard `perfect integrate-and-fire' (PIF) neuron introduced in \cite{Gerstein1964}: in addition to a constant leakage current $f(V) = \beta$, which here we consider as a part of the current bias $I_{DC}$, a reflecting barrier at $V = 0$ is set to avoid a divergent diffusion towards negative membrane potentials. Contrary to the PIF neuron, this makes the VIF neuron capable of having a non-zero (positive) mean firing rate also under subthreshold regimes, i.e., for negative drifts ($\mu<0$). For the VIF neuron, input-output gain function $\Phi(\mu,\sigma)$, eigenfunctions and eigenvalues of the related FP operator have explicit analytical expressions \cite{Fusi1999,Mattia2002,Mattia2004}, which make this model particularly suited for matching theory and simulations. The parameters for the VIF neuron networks used in this work are listed in Tab.~\ref{tab:VIF}. For this model the natural unit of measure for the membrane voltage is the emission threshold $\vthr$.

For the LIF \cite{Tuckwell1988} and the EIF \cite{Fourcaud-Trocme2003} neuron networks, the parameters at drift- (DD) and
noise-dominated (ND) regimes are listed in Tabs.~\ref{tab:LIF} and \ref{tab:EIF}, respectively.

\begin{table}[!ht]
\centering
\begin{tabular}{|m{6cm}|m{1cm}|m{1cm}|}
\hline
\textbf{Parameter} & \textbf{Value (DD)} & \textbf{Value (ND)} \\
\hline
Number of excitatory neurons $N$ & 2000 & 2000\\
\hline
Mean excitatory synapses per neuron $C$ & 100 & 100\\
\hline 
Minimum transmission delay $\delta_{min}$ [ms] & 3 & 3 \\
\hline
External spike rate $C_{ext} \, \nu_{ext}$ [Hz] & 12000 & 12000 \\ 
\hline
Excitatory PSP amplitude due to external spikes $J_{ext}$ [mV] & 0.0804 & 1.76 \\
\hline
Spike rate from external neurons $C_{ext}$ [Hz] & 12000 & 12000 \\
\hline
Current bias $I_{DC}$ [pA] & $-8.46$ & $-223$ \\
\hline
Membrane time constant $\tau_m$ [ms] & 10 & 10 \\
\hline
Refractory period $\tau_{ref}$ [ms] & 1.7 & 1.7 \\
\hline
Spike slope factor $\Delta_T$ [mV] & 3.48 & 3.48 \\
\hline
Firing threshold $\vthr$ [mV] & $-59.9$ & $-59.9$ \\
\hline
Membrane capacitance $C_m$ [pF] & 500 & 500 \\
\hline
Resting potential $V_m$ [mV] & $-65$ & $-65$ \\
\hline
Leak potential $E_L$ [mV] & $-65$ & $-65$ \\
\hline
Reset potential $\vres$ [mV] & $-68$ & $-68$ \\
\hline
Leak conductance $g_L = C_m / \tau_m$ [nS] & 50 & 50 \\
\hline
Excitatory PSP amplitude due to recurrent spikes $J$ [mV] & 0.134 & 0.272 \\
\hline
$V_{peak}$ [mV] & 0 & 0 \\
\hline
Firing rate $\nu*$ [Hz] & 40 & 40 \\
\hline
\end{tabular}
\caption{Parameters of the EIF neuron networks at drift-dominated (DD) and noise-dominated (ND) regimes.}
\label{tab:EIF}
\end{table}

The LIF neuron networks for the Hopf bifurcation test (Fig.~\ref{fig:Hopf}) are similar to the one described in Tab.~\ref{tab:LIF}, but with
a doubled connection probability $C/N=0.1$ and an increased number of neurons, which are now inhibitory. The list of changed parameters is in  Tab.~\ref{tab:HOPF1}. Only two values of synaptic time constants were considered: $\taus = 4$ ms (T4) and $\taus = 16$ ms (T16). Table~\ref{tab:HOPF2} lists the values of $J$ and the corresponding values of $I_{DC}$ set to keep constant the fixed point $\nu^*$, either stable or not.

\begin{table}[!ht]
\centering
\begin{tabular}{|m{6cm}|m{1cm}|m{1cm}|}
\hline \textbf{Parameter} & \textbf{Value (T4)} & \textbf{Value (T16)} \\
\hline
Number of inhibitory neurons $N$ & 20000 & 20000 \\
\hline
Mean inhibitory synapses per neuron $C$ & 2000 & 2000 \\
\hline
Excitatory PSP amplitude due to external spikes $J_{ext}$ [mV] & 0.1 & 0.1 \\
\hline
Membrane time constant $\tau_m$ [ms] & 20 & 20 \\
\hline
Refractory period $\tau_{ref}$ [ms] & 0 & 0 \\
\hline
Firing threshold $\vthr$ [mV] & 20 & 20 \\
\hline
Membrane capacitance $C_m$ [pF] & 500 & 500 \\
\hline
Resting potential $V_E$ [mV] & 0 & 0 \\
\hline
Reset potential $\vres$ [mV] & 10 & 10 \\
\hline
Firing rate $\nu*$ [Hz] & 100 & 40 \\
\hline
\end{tabular}
\caption{Parameters of the LIF neuron networks used for the 
Hopf bifurcation test (Fig.~\ref{fig:Hopf}) when $\taus =\taud = 4~$ms (T4) and $\taus = \taud = 16~$ms (T16).}
\label{tab:HOPF1}
\end{table}

\subsection{Comparison between and characterization of simulated network activities}\label{app:C}

The time-dependent emission rate $\nu(t)$ of the simulated networks was computed as an histogram of the spike emission times of the neurons, with bin size $dt = 0.5 \mathrm{ms}$. The power spectral density $P_\nu(\omega)$ was estimated from the network emission rate by means of the standard Welch method, using a window size of 4096 samples with an overlap of 2048 samples.

\begin{table}[!htb]
\centering
\begin{tabular}{|m{2cm}|m{2cm}|}
\hline \textbf{$J$ [mV]} & \textbf{$I_{DC}$ [pA]}\\
\hline $-0.0025$ & $-2020$ \\ 
\hline $-0.0040$ & $-1990$ \\
\hline $-0.0063$ & $-1940$ \\ 
\hline $-0.0100$ & $-1870$ \\
\hline $-0.0158$ & $-1760$ \\ 
\hline $-0.0250$ & $-1570$ \\ 
\hline $-0.0628$ & $-838$ \\ 
\hline $-0.0995$ & $-134$ \\ 
\hline $-0.158$ & $970$ \\ 
\hline $-0.250$ & $2700$ \\ 
\hline
\end{tabular}
\caption{Set of $J$ and $I_{DC}$ values used in the network of inhibitory LIF neurons described in Tab.~\ref{tab:HOPF1} to produce the results plotted in Fig.~\ref{fig:Hopf}.}
\label{tab:HOPF2}
\end{table}

\end{document}